\documentclass[pra,superscriptaddress,showpacs,aps,10pt]{revtex4-1}

\usepackage{xcolor}
\usepackage{colortbl}
\usepackage{graphicx}
\usepackage{amssymb,amsmath}
\usepackage{subfigure}
\usepackage{enumitem}
\usepackage{bm}
\usepackage{textcomp}
\usepackage{gensymb}
\usepackage{braket}
\usepackage{siunitx}
\usepackage[normalem]{ulem}
\usepackage{float}
\usepackage{overpic}

\usepackage[T1]{fontenc}

\usepackage[%
  bookmarks=true,
  colorlinks,
  linkcolor=blue,
  urlcolor=black,
  citecolor=blue,
  plainpages=false,
  pdfpagelabels,
  final,
  breaklinks=true,
]{hyperref}
\hypersetup{
  pdftitle={Two-Electron Effects Extend High-Harmonic Generation into the keV Regime}, 
  pdfauthor={Isobel McSweeney, Andres Marchisio, Javier Rivera-Dean, Philipp Stammer, Paraskevas Tzallas, Marcelo F. Ciappina and Maciej Lewenstein}
}

\makeatletter
\renewcommand\frontmatter@abstractwidth{\dimexpr\textwidth-1.15in\relax}
\makeatother

\graphicspath{{Figures/}{}}

\usepackage[sort&compress]{natbib}
\begin{document}

\title{Two-Electron Effects Extend High-Harmonic Generation into the keV Regime}

\author{Isobel McSweeney}
\affiliation{ICFO -- Institut de Ciencies Fotoniques, The Barcelona Institute of Science and Technology, 08860 Castelldefels (Barcelona), Spain}

\author{Andrés Marchisio}
\affiliation{Department of Physics, Guangdong Technion - Israel Institute of Technology,
241 Daxue Road, Shantou, Guangdong, China, 515063}
\affiliation{Technion – Israel Institute of Technology, Haifa, 32000, Israel}

\author{Javier Rivera-Dean}
\affiliation{Department of Physics and Astronomy, University College London, Gower Street, London WC1E 6BT, UK}

\author{Philipp Stammer}
\affiliation{ICFO -- Institut de Ciencies Fotoniques, The Barcelona Institute of Science and Technology, 08860 Castelldefels (Barcelona), Spain}
\affiliation{Atominstitut, Technische Universit\"{a}t Wien, 1020 Vienna, Austria}

\author{Paraskevas Tzallas}
\affiliation{Foundation for Research and Technology-Hellas, Institute of Electronic Structure \& Laser, GR-70013 Heraklion (Crete), Greece}
\affiliation{Center for Quantum Science and Technologies (FORTH-QuTech), GR-70013 Heraklion (Crete), Greece}
\affiliation{ELI-ALPS, ELI-Hu Non-Profit Ltd., Dugonics tér 13, H-6720 Szeged, Hungary}

\author{Marcelo F. Ciappina}
\affiliation{Department of Physics, Guangdong Technion - Israel Institute of Technology,
241 Daxue Road, Shantou, Guangdong, China, 515063}
\affiliation{Technion – Israel Institute of Technology, Haifa, 32000, Israel}
\affiliation{Guangdong Provincial Key Laboratory of Materials and Technologies for Energy Conversion,
Guangdong Technion – Israel Institute of Technology, 241 Daxue Road, Shantou, Guangdong, China, 515063}

\author{Maciej Lewenstein}
\email{maciej.lewenstein@icfo.eu}
\affiliation{ICFO -- Institut de Ciencies Fotoniques, The Barcelona Institute of Science and Technology, 08860 Castelldefels (Barcelona), Spain}

\date{\today}

\begin{abstract}

Two-electron processes can generate high harmonics beyond the conventional single-active-electron cutoff. Motivated by recent experimental evidence of an extended secondary plateau in the helium high-harmonic spectrum [S. Wang et al, Optica, (2023); S. Wang et al, In Print in Nature Photon., (2026)], we present a two-electron generalisation of the strong-field approximation. We analyse the resulting expressions using the saddle-point method and determine the extended cutoff. We find good agreement with classical predictions of cutoff scalings of $4.7$ and $5.5$ times the ponderomotive energy, which significantly exceed the established single-electron scaling of 3.17. We calculate high-harmonic spectra generated via a two-electron process in helium atoms driven by an intense few-cycle infrared laser pulse. Our results demonstrate that the harmonic spectrum extends far beyond the water window, reaching photon energies up to $\approx 1.2\,\mathrm{keV}$ in the soft x-ray region. The large spectral bandwidth can support the generation of sub-attosecond soft x-ray pulses, which are of particular interest for probing ultrafast dynamics across matter, including applications in core-level spectroscopy and biological imaging.

$\ $
\noindent
We dedicate this work to the memory of Kenneth C. Kulander, who died in June 2024 at the age of 80, and Joseph H. Eberly, who passed away in April 2025 at the age of 89. 
$\ $
\vspace{7.75pt}
\end{abstract}

\maketitle

\section{Introduction}
\label{sec:1-introduction}

The past three decades have seen remarkable progress in our understanding of nonlinear laser-matter interactions, with high-harmonic generation (HHG) firmly established as a central tool in attoscience research. This has driven a revolution in laser technology, culminating in the $2018$ Nobel Prize in Physics awarded to G\'erard Mourou and Donna Strickland~\cite{Strickland1985, CPA-Nobel-2018}, and, more recently, the $2023$ Nobel Prize awarded to Anne L'Huillier, Pierre Agostini and Ferenc Krausz~\cite{Agostini_Nobel_2024,Krausz_Nobel_2024,Huillier_Nobel_2024}. It is now routinely possible to generate few-cycle femtosecond (fs) laser pulses spanning the visible to mid-infrared regimes~\cite{Maiuri2020, Hemmer2013}, enabling the study of atomic and molecular interactions with coherent electromagnetic radiation at intensities approaching those of atomic binding fields~\cite{Ciappina2017,Krausz2009}. Beyond this, current research efforts are focused on the generation and control of attosecond (as) light pulses~\cite{Varillas2026}. In this context, HHG has facilitated the production of extreme ultraviolet attosecond pulse trains~\cite{Paul2001}, isolated attosecond pulses~\cite{Drescher2001, Chang2012}, and even pulses extending into the soft x-ray regime~\cite{Drescher2001, Silva2015, Cousin2017}. Recently, pulses as short as $19.2\,\mathrm{as}$ were achieved, surpassing the atomic unit of time~\cite{Ardana-Lamas2025}. These coherent pulses~\cite{Pascalcoh, AnneML} are ideal for dynamical spectroscopy, providing insight into the structure of atomic and molecular targets~\cite{Itatani2004, Blaga2012, Pullen2015}, as well as the properties of the driving laser field~\cite{Paulus2001}. Decoding the nonlinear signals requires a comprehensive and, ideally, as analytical as possible theoretical understanding of the underlying processes. This is where the strong-field approximation (SFA) proves valuable~\cite{symphony}. 

HHG is well understood in the single-active-electron (SAE) approximation through the three step model. An electron is liberated from an atom or molecule via tunnel ionisation~\cite{Keldysh1965}, which typically occurs near the maximum of the electric field. The electron accelerates in the oscillating field and recombines with the parent ion, emitting a high frequency photon. This model, also known as the ``simple man's model''~\cite{Corkum1993, Krause1992, Kulander1993}, offers a simple and intuitive description of this phenomenon, in which the electron dynamics are treated classically. However, a more accurate understanding requires a semiclassical framework treating the electron dynamics quantum-mechanically, as reviewed in Ref.~\cite{symphony}. The $1994$ semiclassical SFA formulation by Lewenstein et al.~\cite{Lewenstein1994} was a major milestone in the history of attoscience. It predicts the dynamics of electron quantum paths in intense laser-atom interactions, from which one can obtain the high-harmonic spectrum, the spectral phase distribution, and the maximum harmonic photon energy, beyond which the harmonic amplitudes fall off rapidly.

As a testament to the success of this approach, the field of intense laser-matter interaction, HHG, and attosecond science has largely been built upon the SAE approximation, with multielectron effects receiving comparatively less attention. Several theoretical approaches developed to bridge this gap rely primarily on $S$-matrix theory \`a la W.~Becker~\cite{Kopold2000, Becker2012} and on \textit{ab initio} time-dependent Schr\"odinger equation (TDSE) calculations~\cite{Lappas1996, Chacon2016, Efimov2018}. The SFA formalism was recently applied to non-sequential double ionisation (NSDI)~\cite{symphony}, in particular to the resonant excitation with subsequent ionisation (RESI) and electron impact ionisation (EII) mechanisms. In addition, Koval et al.~\cite{Koval2007} studied multielectron effects in HHG and proposed the non-sequential double recombination (NSDR) mechanism. They introduced an “extended SFA”, which uses Volkov propagators and approximates scattering states as plane waves. NSDR has also been studied by solving the TDSE for both atomic systems~\cite{Feng2015,Feng2015a} and the molecular system $\mathrm{H}_2$~\cite{Hansen2016, Hansen2017, Iravani2018}. Despite the theoretical interest, there was, until recently, a lack of experimental data concerning multielectron effects in HHG.

Recent pioneering work by Wang et al.~\cite{Wang-preprint23, Wang-preprint26} reported the first experimental observation of an HHG spectrum exhibiting signatures of multielectron effects induced by the interaction of strong femtosecond laser fields with helium atoms. The authors observed a significant extension of the harmonic spectrum with photon energies reaching up to $280\,\mathrm{eV}$, exceeding the single-electron cutoff and approaching the water window ($284\text{--}543\,\mathrm{eV}$)~\cite{Takahashi2008}.
Motivated by these findings, in the present work we formulate a fully analytical model \`a la Lewenstein~\cite{Lewenstein1994} for correlated two-electron systems within the SFA framework. We consider the NSDR mechanism, in which sequential double ionisation is followed by simultaneous double recombination. Using the interaction of helium atoms with an intense few-cycle infrared ($800\,\mathrm{nm}$) laser field, we calculate the harmonic spectrum generated by simultaneous double-electron recombination. We observe an extended plateau with sufficient bandwidth to support sub-attosecond pulse generation, reaching a cutoff photon energy of $\approx 1.2\,\mathrm{keV}$.

The paper is organised as follows: In Section \ref{sec:II-background}, we introduce the key assumptions underlying the two-electron SFA. In Section \ref{sec:III-TDSE}, we derive the two-electron dynamical Landau-Dykhne formula for the simultaneous double recombination transition. In Section \ref{sec:IV-Saddle}, we analyse the resulting expression in the saddle-point approximation. Section \ref{sec:V-numerics} discusses numerical results using a short, few-cycle driving pulse, comparing the spectra to the recent experimental observations~\cite{Wang-preprint23, Wang-preprint26}. In addition, we present ultrashort soft x-ray pulses obtained by filtering high harmonics. We conclude in Section \ref{sec:VI-conclusion}.

\section{Formulating the Two-Electron SFA}
\label{sec:II-background}

Neglecting nuclear motion, the TDSE provides a complete description of an atomic or molecular system in an intense electromagnetic field. 
Solving the TDSE yields the time evolution of the wavefunction and associated physical observables, allowing for the numerical study of nonlinear strong-field processes such as HHG~\cite{KulanderGaarde1998, TateDiMauro2007, Marcelo2012, Scrinzi2014} and above-threshold ionisation (ATI)~\cite{Kulander1987, Muller1999, Bauer2005classical, Blaga2009}.  
However, performing the full numerical integration of the TDSE over all degrees of freedom is computationally very demanding, if at all possible. 
Moreover, developing a physical interpretation from these numerical results is highly nontrivial, as expected for an {\it ab initio} technique. 
As a result, approximations that provide physical insight continue to play an important role, and this is where the SFA has proven its worth over the years.

\begin{figure}[H]
    \centering
    \includegraphics[width=0.7\linewidth]{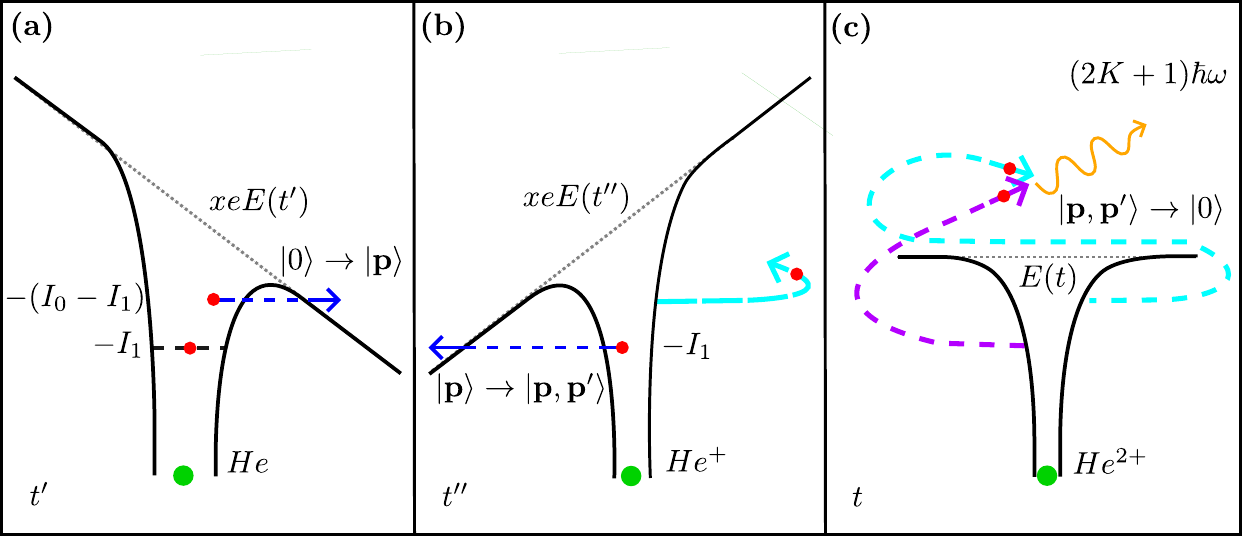}
    \caption{Non-sequential double recombination (NSDR) for a helium atom in a strong laser field. (a) The first electron undergoes tunnel ionisation, transitioning from the ground state $|0\rangle$ to the single-electron continuum $|\textbf{p}\rangle$ at time $t'$. (b) At time $t''$, the second electron tunnels from the more confining effective potential of $\mathrm{He}^+$ into the continuum $|\textbf{p},\textbf{p}'\rangle$ at the subsequent field maximum. (c) The two electrons recombine simultaneously with the parent ion at time $t$, emitting an odd harmonic photon. Note that this is a schematic representation of the electron trajectories.}
    \label{fig:two_electron_diagram}
\end{figure}

Here, we generalise the SFA to analyse two-electron processes in HHG. We consider the NSDR mechanism first proposed in Ref. \cite{Koval2007}. The first electron is ionised via tunnelling from the distorted atomic Coulomb potential and accelerates away from the parent ion under the influence of the laser field. During a subsequent field maximum, the second electron independently tunnels into the continuum. This occurs with a lower probability due to the increased ionisation potential. According to classical analysis~\cite{Wang-preprint23, Wang-preprint26, Koval2007}, this is expected to occur in either the next half-cycle or the following one. The two electrons recombine simultaneously with the parent ion, emitting high-energy radiation. This process is illustrated in Fig. \ref{fig:two_electron_diagram}.

\subsection{Hamiltonian and TDSE}

We consider helium as a two-electron system interacting with a linearly polarised intense laser field within the framework of the two-active-electron (TAE) approximation. Assuming the laser wavelength $\lambda$ is large compared to the Bohr radius $a_0$, the electric field of the laser beam can be considered spatially homogeneous in the region of interest. This is known as the dipole approximation. Relativistic effects that may become important at very high intensities are neglected~\cite{Reiss2018}. The laser electric field is given by:
\begin{equation} \label{Eq:Efield}
 \textbf{E}(t) = E_0\:f(t) \sin(\omega \:t - \phi)\,{\bf e}_{x},
\end{equation}
where $E_0$ is the field amplitude, $\omega = \frac{2\pi c}{\lambda}$ the carrier frequency, $c$ the speed of light, ${f(t)}$ the pulse envelope, $\phi$ the phase and $x$ the polarisation axis.
The dynamics of the two-electron state $| \Psi(t) \rangle$ are governed by the TDSE:
\begin{equation}
i\hbar\frac{ \partial}{ \partial t} | \Psi(t) \rangle=\hat{H} | \Psi(t) \rangle , \label{Eq:SE}
\end{equation}
where the Hamiltonian $\hat{H}$ describing the laser-matter system in the TAE approximation is the sum of two terms:
\begin{equation} \label{Eq:H} 
    \hat{H} = \hat{H_0}+\hat{U} ,
\end{equation}
where $\hat{H}_0$ is the Hamiltonian of the atomic system in the absence of the laser field:
\begin{equation}
    \hat{H}_0=- \frac{\hbar^2\nabla^2_1}{2m} - \frac{\hbar^2\nabla^2_2}{2m} + V({\hat{\textbf{r}}_1, \hat{\textbf{r}}_2}) ,
\end{equation}
with $V({\hat{\textbf{r}}_1, \hat{\textbf{r}}_2})$ the effective TAE potential and $m$ the electron mass. Under the dipole approximation, the interaction term describing the coupling between the atomic system and the laser field can be written in the length gauge~\cite{Bauer2005, Galstyan2016} as:
\begin{equation}
    \hat{U}=-e{\bf E}(t)\cdot (\hat{\bf r}_1+ \hat{\bf r}_2) .
\end{equation}
Note that throughout this analysis, $e$ denotes the electron charge, so that $e<0$.

\subsection{Two-electron SFA \`a la Lewenstein}

As in Ref.~\cite{symphony}, we consider laser fields with low frequency and relatively high intensity, for which the SFA is expected to be most applicable~\cite{Keldysh1965, Faisal1973, Reiss1980, Lewenstein1994,Lewenstein1995, Grochmalicki1986}. This corresponds to the tunnelling regime characterised by a Keldysh parameter $\gamma = \sqrt{I_p/2U_p}<1$, where $I_p$ is the ionisation energy and $U_p=(e^2E_0^2)/(4m\omega^2)$ is the ponderomotive energy, defined as the cycle-averaged kinetic energy gained by an electron in the field.
In this regime, the continuum electron dynamics are dominated by the intense laser field, allowing the effective atomic potential to be treated as a perturbation. 
Nevertheless, rescattering processes play an important role in many phenomena, especially ATI~\cite{Pascal2001} and two-electron ionisation mechanisms, such as EII and RESI~\cite{symphony}. As a result, we make the following assumptions in our formulation of the two-electron analogue of the ``standard SFA''~\cite{Lewenstein1994}:

\begin{enumerate}

\item[(i)] The strong laser field couples the ground state $|0\rangle$ directly to the continuum, while couplings to excited bound states are neglected. The dynamics therefore involve only the ground state and the continuum states $|\textbf{p}\rangle$ and $|\textbf{p}, \textbf{p}'\rangle$. 

\item[(ii)] We shift the energy of the ground state to zero by shifting the Hamiltonian $\hat{H}_0\to\hat{H}_0 + I_0$, where $I_0>0$ is the ionisation energy, given by $I_0=-E_0$, and $E_0$ is the ground state energy of the bound two-electron system: 
\begin{equation} \label{Eq:eigen_0}
\hat{H}_0 |0\rangle = 0|0\rangle.
\end{equation}
This shift in reference energy is equivalent to a unitary phase transformation $|\Psi(t)\rangle\to\exp{(iI_0t/\hbar)}|\Psi(t)\rangle$.
\item[(iii)] 
The amplitude of the ground state $a(t)$ is assumed to be known.

\item[(iv)] 
The continuum states are constructed from the basis of {\it exact} scattering  states, which are eigenstates of the atomic Hamiltonian $\hat{H}_0$. The singly-ionised continuum states $|\textbf{p}\rangle$ are characterised by the outgoing momentum ${\bf p}$ and satisfy:
\begin{equation} \label{Eq:eigen_p}
\hat{H}_0 |\textbf{p}\rangle = \left[\frac{1}{2m}\textbf{p}^2 + I_0-I_1\right]|\textbf{p}\rangle,
\end{equation}
where $I_1$ denotes the binding energy of the single remaining bound electron. The double-continuum states $|\textbf{p}, \textbf{p}'\rangle$, characterised by the outgoing momenta ${\bf p}$, ${\bf p'}$, satisfy:
\begin{equation} \label{Eq:eigen_ps}
\hat{H}_0 |\textbf{p}, \textbf{p}'\rangle = \left[\frac{1}{2m}\textbf{p}^2 + \frac{1}{2m}(\textbf{p}')^2+ I_0\right]|\textbf{p}, \textbf{p}'\rangle.
\end{equation}

\end{enumerate}

We define the following matrix elements:
\begin{enumerate}
\item The dipole matrix element describing the transition from the ground state to the single-electron continuum:
\begin{equation} \label{Eq:dv}
e \langle \textbf{p} |\hat{\textbf{r}}_1 + \hat{\textbf{r}}_2|0 \rangle
=
\textbf{d}_0(\textbf{p}),
\end{equation}
where the subscript $0$ indicates the transition originates from the state $|0\rangle$, whose ground state energy $-I_0$. 

\item The continuum-continuum matrix elements coupling the states ${\bf p}$ and ${\bf p}'$:
\begin{equation} \label{Eq:dv2}
e \langle \textbf{p}' |\hat{\textbf{r}}_1 + \hat{\textbf{r}}_2|\textbf{p} \rangle
=
\textbf{G}(\textbf{p, p}').
\end{equation}
These matrix elements can be decomposed into their most singular part, proportional to $i \hbar {\nabla}_{\bf p}\delta({\bf p}-{\bf p}')$, and the ``rest''~\cite{Grochmalicki1986, Lewenstein1994, Lewenstein1995}. The ``rest'' is then treated in a perturbative manner~\cite{Lewenstein1995}.  

\item The dipole matrix elements describing the simultaneous transition from the ground state to the two-electron continuum (and, conversely, simultaneous double recombination):
\begin{equation} \label{Eq:dv3}
e \langle \textbf{p},  \textbf{p}' |\hat{\textbf{r}}_1 + \hat{\textbf{r}}_2|0 \rangle
=
\textbf{d}(\textbf{p}, \textbf{p}' ).
\end{equation}
Although simultaneous tunnelling of two electrons to the continuum is unlikely in atomic systems~\cite{Hansen2016}, the inverse process has been experimentally observed~\cite{Wang-preprint23, Wang-preprint26}.

\item The dipole transition matrix elements describing tunnelling from states with a single electron in the continuum to the two-electron continuum:
\begin{equation} \label{Eq:dv4}
e \langle \textbf{p},  \textbf{p}' |\hat{\textbf{r}}_1 + \hat{\textbf{r}}_2|\textbf{p}'' \rangle
=
\textbf{G}(\textbf{p}, \textbf{p}'; \textbf{p}'' ),
\end{equation}
where $|\textbf{p}''\rangle$ is a single-electron continuum eigenstate of $H_0$, as defined in Eq. \eqref{Eq:eigen_p}.

\item Finally, we require the two-electron continuum-continuum matrix elements:
\begin{equation} \label{Eq:dv5}
    e\langle\textbf{p}, \textbf{p}'| \hat{\textbf{r}}_1 + \hat{\textbf{r}}_2| \textbf{p}'' , \textbf{p}''' \rangle = \textbf{G}(\textbf{p}, \textbf{p}'; \textbf{p}'', \textbf{p}''' ),
\end{equation}
where $|\textbf{p}'' , \textbf{p}'''\rangle$ is a double-electron continuum eigenstate of $H_0$, as defined in Eq. \eqref{Eq:eigen_ps}.

\end{enumerate}

\vspace{0.5cm}\noindent
In the spirit of previous studies~\cite{Lewenstein1994,Lewenstein1995,symphony}, we now expand upon and clarify these points below:

\begin{itemize}
\item
According to statement (i), the electronic state $|\Psi(t)\rangle$ is a coherent superposition of the ground state $|0 \rangle$,  the single-electron continuum states, $|\textbf{p}\rangle$, and the two-electron continuum states, $|\textbf{p}, \textbf{p}'\rangle$:
\begin{equation} \label{Eq:psi}
|\Psi(t) \rangle= a(t) |0 \rangle \: + \: \int{\textit{d}^3 \textbf{p} \:  \textit{b}( \textbf{p},t) |\textbf{p}\rangle} \: +
 \: \int{\textit{d}^3 \textbf{p} \int{\textit{d}^3 \textbf{p}' \:  \textit{c}( \textbf{p}, \textbf{p}',t) |\textbf{p} , \textbf{p}'\rangle}}  .
\end{equation}
The transition amplitude to the single-electron continuum states is denoted by  $\textit{b}(\textbf{p},t)$ and depends on both the kinetic momentum of the outgoing electron and the laser pulse. 
The transition amplitude to the two-electron continuum states is denoted by 
$\textit{c}( \textbf{p}, \textbf{p}',t)$. 

Note that other bound states can be included as additional terms in Eq.~\eqref{Eq:psi} (cf.\ Refs.~\cite{Ivanov2014, PerezHernandez2009, Sanpera1996}).

\item
Several methods are available to evaluate or estimate the ground state amplitude $a(t)$, depending on the parameter regime. I) $a(t)$ can be determined by solving the TDSE for the target system {\it ab initio}. II) $a(t)$ can be calculated with any ``cheap'' approximate method, such as classical phase-space averaging~\cite{Berman2018}. III) A widely used method is to evaluate $a(t)$ analytically using ionisation rates from Ammosov-Delone-Krainov (ADK) theory~\cite{Ammosov1986}. IV) For short pulses, or for long pulses of sufficiently low intensity, ground-state depletion can be neglected and the phase varies slowly, so that $a(t)\simeq 1$. V) $a(t)$ can be calculated self-consistently within the SFA, as discussed in Ref.~\cite{Lewenstein1994}.

\item
The continuum-continuum matrix elements describing transitions between single-electron continuum states can be written in the general form:
\begin{equation} \label{Eq:c-c}
\textbf{G}(\textbf{p, p}') = ie \hbar {\nabla}_{\bf p}\delta({\bf p}-{\bf p}') + \hbar {\bf g}({\bf p},{\bf p}'),
\end{equation}
where the first term is the most singular contribution and the second term represents all other contributions, including Coulomb effects arising from the atomic potential. These are responsible for rescattering and play an important role in ATI, EII and RESI. Typically the strongest singularity contained in $\hbar{\bf g}({\bf p},{\bf p}')$ is associated with the on-energy-shell gradient of the Dirac delta function $\delta({\bf p}^2-({\bf p}')^2)$, reflecting the coupling of continuum states with equal energies. 
The above expression holds for both short-range effective potentials (e.g. model atoms and negative ions) and Coulomb-like potentials.
Note that the use of {\it exact} scattering states incorporates both the short-range effects and any long-range Coulomb effects of the effective potential directly into the dipole matrix element $\textbf{d}_0(\textbf{p})$, as well as into the rescattering continuum-continuum matrix elements $\textbf{G}(\textbf{p}, \textbf{p}'; \textbf{p}'' )$ and $\textbf{G}(\textbf{p}, \textbf{p}'; \textbf{p}'', \textbf{p}''' )$.
Many previous studies assumed Volkov (plane-wave) continuum states in the SFA (e.g. Ref.~\cite{Koval2007,Ivanov2014}) rather than {\it exact} scattering states. However, this is an additional approximation and there is a danger in overlooking the underlying assumptions, especially in the case of molecules and other extended targets (see Ref.~\cite{symphony} for a detailed discussion).

\item
In a similar spirit, we determine the leading part of the matrix elements 
$\textbf{G}(\textbf{p}, \textbf{p}'; \textbf{p}'' )$. These correspond to processes expected to be weak, so we retain only the leading singular contribution. We make the following assumptions in our approximation: 
\begin{enumerate}
    \renewcommand{\labelenumi}{\roman{enumi})}
    \item We approximate the wavefunctions of the involved states as symmetrised product states.

    \item The continuum states are approximated by plane waves.

    \item The bound state, corresponding to a single electron with binding energy $I_1$, is described by the wavefunction $\Psi_1(\textbf{x})$.

    \item We carefully respect the fact that $\Psi_1(\textbf{x})$ is orthogonal to the states in the continuum.
    
\end{enumerate}

In effect, we obtain:
\begin{equation} \label{Eq:c-c2}
\textbf{G}(\textbf{p}, \textbf{p}'; \textbf{p}'' )\simeq \frac{1}{2}\left[\delta(\textbf{p}'- \textbf{p}'') \textbf{d}_1(\textbf{p}) + \delta(\textbf{p}- \textbf{p}'') \textbf{d}_1(\textbf{p}')\right],
\end{equation}
where $\textbf{d}_1(\textbf{p})$ is the dipole transition matrix element coupling $\Psi_1(\textbf{x})$ to the continuum.

\item 
Finally, we consider the two-electron continuum-continuum matrix elements. Again, we retain only the most singular contribution:
\begin{equation}
    \begin{aligned}
        \textbf{G}(\textbf{p}, \textbf{p}'; \textbf{p}'',\textbf{p}''' ) \simeq& \frac{1}{2}\left[ 
ie \hbar {\nabla}_{\bf p}\delta({\bf p}-{\bf p}'')\delta(\textbf{p}'- \textbf{p}''') 
+ ie \hbar {\nabla}_{\bf p}\delta({\bf p}-{\bf p}''')\delta(\textbf{p}'- \textbf{p}'') \right. \\ 
+&\left. ie \hbar {\nabla}_{{\bf p}'}\delta(\textbf{p}'-\textbf{p}'')\delta(\textbf{p}- \textbf{p}''')  + ie \hbar {\nabla}_{{\bf p}'}\delta({\bf p}'-{\bf p}''')\delta(\bf{p}- \bf{p}'')\right].
    \end{aligned}
    \label{Eq:c2-c2}
\end{equation}
\end{itemize}

\section{Solving the TDSE}
\label{sec:III-TDSE}

\subsection{TDSE in the SFA Framework}

The main task of this subsection is to derive a
general expression for the transition amplitudes  $\textit{b}({\bf p},t)$ and $\textit{c}({\bf p}, {\bf p}', t)$, which will then allow us to compute HHG spectra.  After some algebra, the time derivative
of the ground state amplitude, $a(t)$, reads:
\begin{equation} \label{Eq:da/dt}
\dot{a}(t) 
 = 
\frac{i}{\hbar}
\int d^3{\bf p} \: 
\textbf{E}(t)\cdot \textbf{d}_0^*\mathopen{}\left( \textbf{p}\right)\mathclose{} \:
b(\textbf{p},t) + \frac{i}{\hbar}
\int d^3{\bf p}\,\int d^3{\bf p}' \: 
\textbf{E}(t)\cdot \textbf{d}^*\mathopen{}\left( \textbf{p}, \textbf{p}'\right)\mathclose{} \:
c(\textbf{p}, \textbf{p}', t).
\end{equation}
The second term in Eq.~\eqref{Eq:da/dt} describes simultaneous transitions to the double continuum, which are considered negligible. However, it plays an important role in the double recombination processes. The time derivative of the single-electron continuum amplitude is given by:

\begin{equation}
    \begin{aligned}
        \dot{b}( \textbf{p},t) 
 &=
-\frac{i}{\hbar}\left(\frac{\textbf{p}^2}{2m}
+ I_0-I_1\right)b( \textbf{p},t) 
+\frac{i}{\hbar}\textbf{E}(t) \cdot {\bf d}_0({\bf p})a(t)
-\,e{\bf E}(t)\cdot\nabla_{\bf p} b({\bf p},t)   \\
&+ i \textbf{E}(t)\cdot \int{\textit{d}^3 \textbf{p}^{\prime}\: {\bf g}( \textbf{p}, \textbf{p}^{\prime})} b( \textbf{p}^{\prime},t) + \frac{i}{\hbar}\textbf{E}(t) \cdot \int{\textit{d}^3 \textbf{p}'} {\bf d}_1^*({\bf p}') c(\textbf{p}, \textbf{p}', t) .
    \end{aligned}
    \label{Eq:db/dt}
\end{equation}

The first term in Eq.~\eqref{Eq:db/dt} represents the free phase evolution of the electron in the absence of the oscillating laser field. 
The second term describes the bound-free transition from the ground state to the single-electron continuum.
The two subsequent terms describe continuum-continuum transitions: $\nabla_{\bf p} b({\bf p},t)$, which accounts for propagation without the influence of the scattering centre, and the term involving the core potential, 
$\int{\textit{d}^3 e\textbf{p}^{\prime}\: b( \textbf{p}^{\prime},t){\bf g}({\bf p},{\bf p}')}$. 
Here ${\bf g}({\bf p},{\bf p}')$ denotes the rescattering transition matrix element, where the core potential plays an essential role. While these terms are essential for rescattering processes, they are not relevant for HHG and are thus neglected in the present paper. The final term describes transitions from the single to double continuum.

Finally, we consider the amplitude in the two-electron continuum:  
\begin{equation}
    \begin{aligned}
        \dot{c}( \textbf{p}, \textbf{p}', t) 
 &=
-\frac{i}{\hbar}\left(\frac{\textbf{p}^2}{2m} + \frac{(\textbf{p}')^2}{2m} + I_0
\right)c( \textbf{p}, \textbf{p}', t) 
-\,e{\bf E}(t)\cdot\nabla_{\bf p} c({\bf p},{\bf p}', t) - \,e{\bf E}(t)\cdot\nabla_{{\bf p}'} c({\bf p},{\bf p}', t)  \\
&+\frac{i}{\hbar}\textbf{E}(t) \cdot {\bf d}({\bf p}, {\bf p}')a(t) + \frac{i}{2 \hbar}\textbf{E}(t) \cdot {\bf d}_1({\bf p}) b(\textbf{p}', t) +  \frac{i}{2 \hbar}\textbf{E}(t) \cdot {\bf d}_1({\bf p}') b(\textbf{p}, t) .
    \end{aligned}
    \label{Eq:dc/dt}
\end{equation}

The first term in Eq.~\eqref{Eq:dc/dt} denotes free propagation of the two electrons in the field-free continuum. The second and third terms
describe the dominant (singular) effects of the laser field on the two-electron propagation. The fourth term
represents simultaneous double ionisation, where both electrons transition from the ground state to the two-electron continuum. This is typically very small and can be neglected. The inverse process, described by the conjugate matrix element, corresponds to simultaneous double recombination. This is the key process investigated in this study. The final two terms describe the transition from the single-electron continuum to the two-electron continuum, which is an essential step in the discussed mechanism. All the formulae above account for symmetry under exchange of the two-electron momenta.

\subsection{Single-electron amplitude in the continuum}
Consider the process in which a single electron is transferred to the continuum (typically via tunnel ionisation). 
This is described by the direct photoelectron transition amplitude $b({\bf p},t)$. 
Here we neglect the rescattering processes~\cite{Lewenstein1995}, as well as the influence of the weak two-electron effects, that is, we disregard the last two terms in Eq.~\eqref{Eq:db/dt}. 
This is what we refer to as the zeroth-order solution:
\begin{equation}
\dot{b}( \textbf{p},t)  
=
-\frac{i}{\hbar}\left(\frac{\textbf{p}^2}{2m}+ I_0-I_1 \right)b( \textbf{p},t)  
+ \frac{\textit{i}}{\hbar}\: \textbf{E}(t) \cdot  \textbf{d}_0( \textbf{p})a(t) - e\textbf{E}(t) \cdot \nabla_{\bf p}b( \textbf{p},t).
\end{equation}

The above equation is a first-order inhomogeneous differential equation, which can be solved by conventional integration methods (see e.g.\ Ref.~\cite{Lev}). The solution is:
\begin{equation}
    \begin{aligned}
        b( \textbf{p},t) 
= & 
\frac{i}{\hbar} \int_0^t{\textit{d} \textit{t}^{\prime}\:\textbf{E}(t^{\prime})}\cdot \textbf{d}_0\left( \textbf{p}+e\textbf{A}(t)/c-e\textbf{A}(t^{\prime})/c\right)
\\
&\times 
\exp\mathopen{}\left(
  -\textit{i} \:
  \int_{t^{\prime}}^{t} d\tilde{t}
  \left[
 \frac{1}{2m} (\textbf{p}+e\textbf{A}(t)/c-e\textbf{A}(\tilde{t})/c)^2
 +I_0-I_1
 \right]/\hbar
 \right)
a({t}^{\prime}). 
    \end{aligned}
    \label{Eq:b0_long}
\end{equation}

Here, we assume the electron appears in the
continuum with kinetic momentum ${\bf p}(t')={\bf p}+e{\bf
A}(t)/c-e{\bf A}(t')/c$ at time $t'$, where {\bf p}~is the final
kinetic momentum and $\textbf{A}(t) =-c\int^{t}{
\textbf{E}(t^{\prime})dt^{\prime}}$ is the vector potential of the
electromagnetic field, with $c$ the speed of light.  
In addition, Eq.~(\ref{Eq:b0_long}) can be simplified by introducing the canonical momentum $\textbf{p}_{c}$, defined by $\textbf{p}_{c} = {\bf p} + e{\bf A}(t)/c$. From this point on we will take $\textbf{p}$ to denote canonical momentum, so the amplitude in the single-electron continuum reduces to the expression in Ref.~\cite{Lewenstein1994}:
\begin{equation} \label{Eq:b_0}
b( \textbf{p},  t) = 
\frac{i}{\hbar} \int_0^t{\textit{d} \textit{t}'\:
\textbf{E}(t')}\cdot \textbf{d}_0\left( \textbf{p}-e\textbf{A}(t')/c\right)
\:
\exp\mathopen{}\left(
  -\textit{i} \:
  \int_{t'}^t{d{\tilde t}}\,
  \left[
  \frac{1}{2m}(\textbf{p}-e\textbf{A}({\tilde t})/c)^2
  +I_0-I_1
  \right]/\hbar
  \right)a(t').
\end{equation}
This expression (Eq.~\eqref{Eq:b_0}) is interpreted as the sum over all possible paths in which the electron is ionised at time $t'$ and later detected at time $t$~\cite{Pascal2001}. 
The argument of the integral corresponds to the instantaneous transition probability amplitude of an electron when it appears in the continuum at time $t'$, and the exponential phase factor, which is identified with the quasi-classical action defining a possible electron trajectory from the ``birth'' time $t'$ until the ``detection'' time $t$~\cite{Lewenstein1995}:

 \begin{equation}
{S}({\bf p},t^{\prime},t) 
= 
\int_{t^{\prime}}^{t}d{\tilde t}
\left[
\frac{1}{2m}({\bf p}-e\textbf{A}({\tilde t})/c)^2
+I_0-I_1
\right].
\end{equation}

Since our goal is to derive the final transition amplitude $b({\bf p},t)$, the detection time $t$ is fixed at the end of the laser pulse, where the pulse envelope vanishes. Therefore, the oscillating electric field satisfies ${\bf E}(0) = {\bf E}(t_{\rm F}) = {\bf 0}$. By construction, the vector potential satisfies the same boundary conditions. In numerical calculations, we define the laser pulse envelope as $f(t)=\sin^2((\omega t)/(2n_c))$, where $n_c$ denotes the total number of cycles.

\subsection{Time-dependent dipole moment in the SAE approximation} 
Finally, the HHG spectrum is obtained from the time-dependent dipole moment. 
Transitions in the continuum are neglected, as is customary within the SFA framework, so the dominant contribution comes from the zeroth-order solution of the SFA equations in the SAE approximation (Eq.~\eqref{Eq:b_0}). 
The expectation value of the dipole moment $\langle {\bf r}(t)\rangle_{1e}$ is then given by the dynamical {\it Landau-Dykhne} formula, as in Refs.~\cite{Lewenstein1994,symphony}:
\begin{equation}
    \begin{aligned}
        e\langle {\bf r}(t)\rangle_{1e} 
& =
\mathrm{Re}\mathopen{}\left[
\frac{2i}{\hbar} 
\int_0^t\textit{d} 
\textit{t}^{\prime} 
\int d^3{\bf p} \:
a^*(t) \,
\textbf{d}_0^*\mathopen{}\left( \textbf{p}-e\textbf{A}(t)/c\right)\mathclose{}
\ 
\textbf{E}(t^{\prime})\cdot \textbf{d}_0\left( \textbf{p}-e\textbf{A}(t^{\prime})/c\right)
a(t^{\prime})
\right. \\ & \left. \times 
\exp\mathopen{}\left(
  -\textit{i} {S}({\bf p},t^\prime,t) / \hbar
  \right)
\right]
,
    \end{aligned}
    \label{Eq:HHG-td-dipole}
\end{equation}

where the single-electron quasi-classical action is given by:

 \begin{equation}
{S}({\bf p},t^\prime,t) 
= 
\int_{t'}^{t}d{\tilde t}
\left[
\frac{1}{2m}({\bf p}-e\textbf{A}({\tilde t})/c)^2 + 
I_0 - I_1
\right].
\end{equation}

Computing the Fourier transform yields the harmonic spectrum, which can be directly compared with experimental results:

\begin{equation}
    \begin{aligned}
        \tilde{\bf r}_{1e}(\Omega)
&= 
\frac{1}{2\pi}\int_{-\infty}^\infty dt \langle {\bf r}(t)\rangle_{1e} \exp{(-i\Omega t)}  
    \end{aligned}
    \label{Eq:HHG-fd-dipole}
\end{equation}
where $\Omega$ is the frequency of the emitted harmonic. This gives rise to the standard HHG spectrum with the SAE cutoff of $3.17U_p+I_p$, as established in Ref.~\cite{Lewenstein1994}.

\subsection{Two-electron amplitude in the double continuum}

We now extend the above analysis to the two-electron case to derive the amplitude in the double continuum. Simultaneous double ionisation is very unlikely to occur, so this pathway was neglected: 
\begin{equation}
    \begin{aligned}
        c(\textbf{p},  \textbf{p}',t) 
&= 
\frac{1}{2}\left(\frac{i}{\hbar}\right)^2 \int_0^t{\textit{d} \textit{t}'}\:{\int_{t'}^t{\textit{d} \textit{t}''}\:
\textbf{E}(t'')}\cdot \Bigl\{ \left[ 
\textbf{E}(t')\cdot\textbf{d}_0\left( \textbf{p} - e\textbf{A}(t')/c \right)\right] \textbf{d}_1\left( \textbf{p}'-e\textbf{A}(t'')/c \right)
\; \exp\mathopen{}\left(
  -\textit{i} \:
  {S}({\bf p},t',t'') /\hbar
  \right)   \\& + \left[ 
\textbf{E}(t')\cdot\textbf{d}_0\left( \textbf{p}' - e\textbf{A}(t')/c \right)\right] \textbf{d}_1\left( \textbf{p}-e\textbf{A}(t'')/c \right)
\; \exp\mathopen{}\left(
  -\textit{i} \:
  {S}({\bf p}',t',t'') /\hbar
  \right)
  \Bigr\}
  \; a(t') \;
  \exp\mathopen{}\left(
  -\textit{i} \:
  {S}({\bf p}, {\bf p}',t'',t) /\hbar
  \right) ,
    \end{aligned}
    \label{Eq:c}
\end{equation}

where the quasi-classical action of the two-electron state in the time interval $[t'',t]$ is:
 \begin{equation}
{S}({\bf p}, {\bf p}', t'',t) 
= 
\int_{t''}^{t}d{\tilde t}
\left[
\frac{1}{2m}({\bf p}-e\textbf{A}({\tilde t})/c)^2 + \frac{1}{2m}({\bf p}'-e\textbf{A}({\tilde t})/c)^2
+I_0
\right], 
\end{equation}
while the quasi-classical action with the single electron in the continuum is:
 \begin{equation}
{S}({\bf p}, t',t'') = 
\int_{t'}^{t''}d{\tilde t}
\left[
\frac{1}{2m}({\bf p}-e\textbf{A}({\tilde t})/c)^2 +I_0 - I_1
\right]. 
\end{equation}
Eq.~\eqref{Eq:c} can be understood as the sum over all possible pathways in which one electron is ionised at time $t'$ and propagates in the continuum until a second electron is ionised at time $t''$. Both electrons subsequently evolve in the continuum until detection time $t$. Note that the expression consists of two terms to account for the symmetrisation with respect to the exchange of momenta ${\bf p} \leftrightarrow {\bf p}'$.

\subsection{Time-dependent dipole moment in the TAE approximation} 

We require the time-dependent dipole moment associated with transitions between the double-continuum state and the ground state to obtain the two-electron HHG spectrum via a Fourier transform.

\begin{equation}
    \begin{aligned}
        e\langle {\bf r}(t)\rangle_{2e} 
& =
\mathrm{Re}\mathopen{}\left[(\frac{\textit{i}}{\hbar})^2 
\int_0^{t}\textit{d} 
\textit{t}' \: \int_{t'}^{t}\textit{d} 
\textit{t}''
\int d^3{\bf p} \: \int d^3{\bf p}'\;
a^*(t) \,
\textbf{d}^*\mathopen{}\left( \textbf{p}-e\textbf{A}(t)/c, \textbf{p}'-e\textbf{A}(t)/c\right)\mathclose{}
\right. \\ & \left. 
\times \textbf{E}(t'')\cdot \Bigl\{ \left[ 
\textbf{E}(t')\cdot\textbf{d}_0\left( \textbf{p} - e\textbf{A}(t')/c \right)\right] \textbf{d}_1\left( \textbf{p}'-e\textbf{A}(t'')/c \right)
\; \exp\mathopen{}\left(
  -\textit{i} \:
  {S}({\bf p},t',t'') /\hbar
  \right) 
\right. \\ & \left. 
+ \left[ 
\textbf{E}(t')\cdot\textbf{d}_0\left( \textbf{p}' - e\textbf{A}(t')/c \right)\right] \textbf{d}_1\left( \textbf{p}-e\textbf{A}(t'')/c \right) \; 
\exp\mathopen{}\left(
  -\textit{i} \:
  {S}({\bf p}',t',t'') /\hbar
  \right)
  \Bigr\}
  \right. \\ & \left.
  \times a(t') \;
  \exp\mathopen{}\left(
  -\textit{i} \:
  {S}({\bf p}, {\bf p}',t'',t) /\hbar
  \right) \right]  .
    \end{aligned}
    \label{Eq:HHG-td-TEA}
\end{equation}

This quantity describes the simultaneous double recombination transition, and the remainder of this paper is devoted to its analysis. It is expected to reproduce the cutoff laws observed in Ref.~\cite{Wang-preprint23, Wang-preprint26}.

\section{Saddle-point analysis}
\label{sec:IV-Saddle}

The resulting expression (Eq.~\eqref{Eq:HHG-td-TEA}) contains four integrals, with a fifth introduced by taking the Fourier transform. 
We now use the saddle-point method to solve the momentum integrals analytically for a linearly polarised electric field. 
This approach is necessary because the phase factors associated with the quasi-classical action are highly oscillatory. 
As a result, the integrand exhibits strong cancellations, which poses a significant challenge for direct numerical integration by demanding a high degree of precision.
Guided by the original SFA reference, Ref.~\cite{Lewenstein1994}, we will take the following steps: i) identify the total quasi-classical action characterising two-electron HHG; ii) derive saddle-point equations from this action; iii) analyse their solutions for a constant-envelope electric field, $E = E_0\cos{(\omega t)}$. Full details of the analysis can be found in the appendix.

\subsection{Saddle-point analysis for two-electron HHG}

Eq.~\eqref{Eq:HHG-td-TEA} consists of two terms, symmetric with respect to exchange ${\bf p} \leftrightarrow {\bf p}'$. We will apply our analysis to each term separately as the results are equivalent. The total (effective) action of the first term in Eq.~\eqref{Eq:HHG-td-TEA} is given by: 
\begin{equation}
    \begin{aligned}
        {S}_{\rm eff}({\bf p}, {\bf p}', t', t'',t) =
&\int_{t''}^{t}d{\tilde t}
\left[
\frac{1}{2m}({\bf p}-e\textbf{A}({\tilde t})/c)^2 + \frac{1}{2m}({\bf p}'-e\textbf{A}({\tilde t})/c)^2
+I_0
\right] \\& + \int_{t'}^{t''}d{\tilde t}
\left[
\frac{1}{2m}({\bf p}-e\textbf{A}({\tilde t})/c)^2 +I_0 - I_1
\right] - (2K+1)\hbar\omega t, 
    \end{aligned}
    \label{Eq:s_eff}
\end{equation}
where we have subtracted the anticipated energy of the emitted $(2K+1)$-th harmonic photon. This is introduced naturally by taking the Fourier transform. Since the expression breaks the symmetry, we will associate with ${\bf p}$ (${\bf p}'$) an electron at the location ${\bf x}_1$
(${\bf x}_2$). We will restore the symmetry later, when necessary. The total (effective) action depends on the two canonical momenta and on the three times. The saddle-point equations for momenta indicate that the two electrons tunnel out at $t'$ and $t''$, respectively, and later recombine with the parent ion at time $t$:
\begin{align}
&\int_{t'}^{t}d{\tilde t}
\left[
\frac{1}{m}({\bf p}-e\textbf{A}({\tilde t})/c)\right] = {\bf x}_1(t)  - {\bf x}_1(t') = 0, \label{Eq:x1}\\
&\int_{t''}^{t}d{\tilde t}
\left[
\frac{1}{m}({\bf p}'-e\textbf{A}({\tilde t})/c)\right] = {\bf x}_2(t)  - {\bf x}_2(t'') = 0. \label{Eq:x2}
\end{align}

The saddle-point equations with respect to the times encode energy conservation laws. These include both the standard classical conditions and quantum-corrected versions associated with classically-forbidden tunnelling: 

\begin{align}
&\frac{1}{2m}({\bf p}-e\textbf{A}(t')/c)^2 + I_0 - I_1= 0,
\label{Eq:saddle_t1}
\\
&\frac{1}{2m}({\bf p'}-e\textbf{A}(t'')/c)^2 + I_1 = 0 ,
\label{Eq:saddle_t2}
\\
\frac{1}{2m}({\bf p}-e\textbf{A}(t)/c)^2&+ \frac{1}{2m}({\bf p}'-e\textbf{A}(t)/c)^2 + I_0 = (2K+1)\hbar\omega .
\label{Eq:saddle_t3}
\end{align}

The first two equations (Eqs.~\eqref{Eq:saddle_t1} and~\eqref{Eq:saddle_t2}) describe the sequential tunnelling processes at $t'$ and $t''$. The electron kinetic energy must be negative at these times, so the solutions of the saddle-point equations are complex. The final equation (Eq.~\eqref{Eq:saddle_t3}) describes ``conventional'' energy conservation in the recombination process.

\subsection{Solving saddle-point equations for two-electron HHG}
We proceed with the goal of keeping the analysis as analogous as possible to that of Ref.~\cite{Lewenstein1994}. 
For this purpose, we introduce dimensionless units by rescaling $t \to \omega t$. 
We also introduce return times for the two electrons: $t-t'=\tau$ and $t-t''=\tau'$. 
In addition, we express all relevant energies in terms of the ponderomotive energy $U_p$, ionisation potentials $I_0$ and $I_1$, and $\hbar\omega$. We now solve the saddle-point equations for momenta (Eqs. \eqref{Eq:x1} and \eqref{Eq:x2}) to obtain the stationary values:
\begin{align}
p_{\rm st}(t, \tau)&= \frac{eE_0}{\tau}[\cos(t) -\cos(t-\tau)], \label{Eq:saddle_p1}\\
p_{\rm st}(t, \tau^\prime)&= \frac{eE_0}{\tau^\prime}[\cos(t) -\cos(t-\tau^\prime)]. \label{Eq:saddle_p2} 
\end{align}
We then  stack the solutions into the expression for the action (Eq. \eqref{Eq:s_eff}):
\begin{equation}
    \begin{aligned}
        S_{\rm st}(t, \tau, \tau^\prime) &= (I_0-I_1+ U_p)\tau - \frac{2U_p}{\tau} [1-\cos(\tau)] - U_pC(\tau)\cos(2t-\tau)  \\
&+ (I_1 + U_p)\tau^\prime - \frac{2U_p}{\tau^\prime} [1-\cos(\tau^\prime)] - U_pC(\tau')\cos(2t-\tau^\prime) - (2K+1) t,
    \end{aligned}
\end{equation}
where
\begin{equation}
C(\tau) = \sin(\tau)-\frac{4\sin^2(\tau/2)}{\tau} .
\end{equation}
Note that here $U_p$ is dimensionless, expressed in units of $\hbar\omega$. First, we need to maximise $S_{\rm st}(t, \tau, \tau')$ with respect to $t$. This leads to the equation: 
\begin{equation} \label{Eq:s_max1}
2U_pC(\tau)\sin(2t-\tau) + 2U_pC(\tau')\sin(2t-\tau^\prime) - (2K+1) = 0 ,
\end{equation}
which can be transformed into the form:
\begin{equation} \label{Eq:s_max2}
A(\tau, \tau')\sin\left(2t-\tau-\phi(\tau, \tau')\right)= B,
\end{equation}
where the analytic expression of $\phi((\tau, \tau')$ can be obtained, but is not relevant to the present discussion. $B$ is the photon energy in units of $U_p$, and $A(\tau, \tau')$ (not to be confused with the vector potential $\textbf{A}(t)$) is given by:
\begin{equation} \label{Eq:A}
A(\tau, \tau') = \sqrt{\left(2C(\tau) + 2C(\tau')\cos(\tau-\tau')\right)^2 + \left(2C(\tau')\sin(\tau-\tau')\right)^2}.
\end{equation}

\begin{figure}
\centering
\begin{overpic}[width=0.48\linewidth]{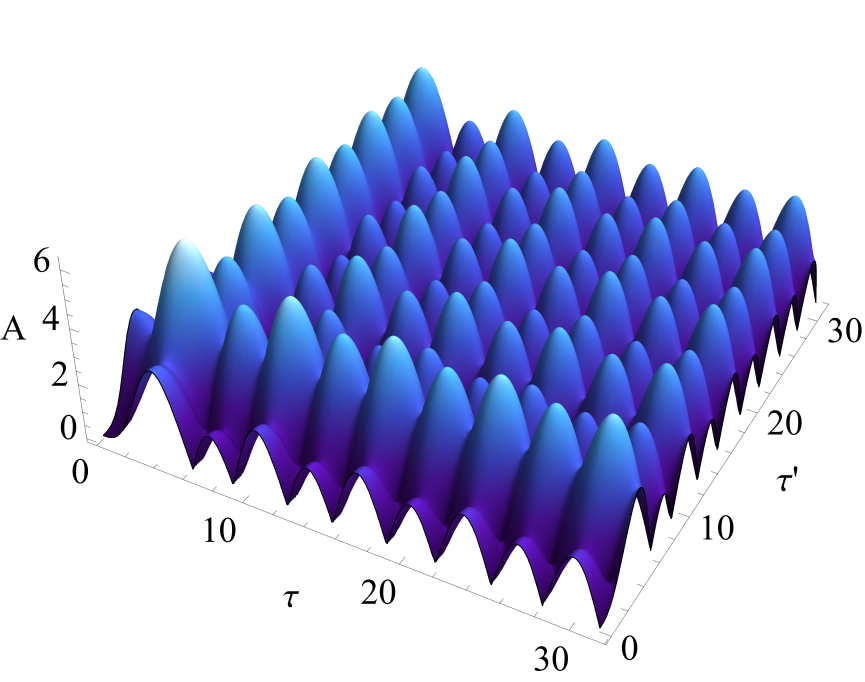}
  \put(3,70){\textbf{(a)}}
\end{overpic}
\hfill
\begin{overpic}[width=0.48\linewidth]{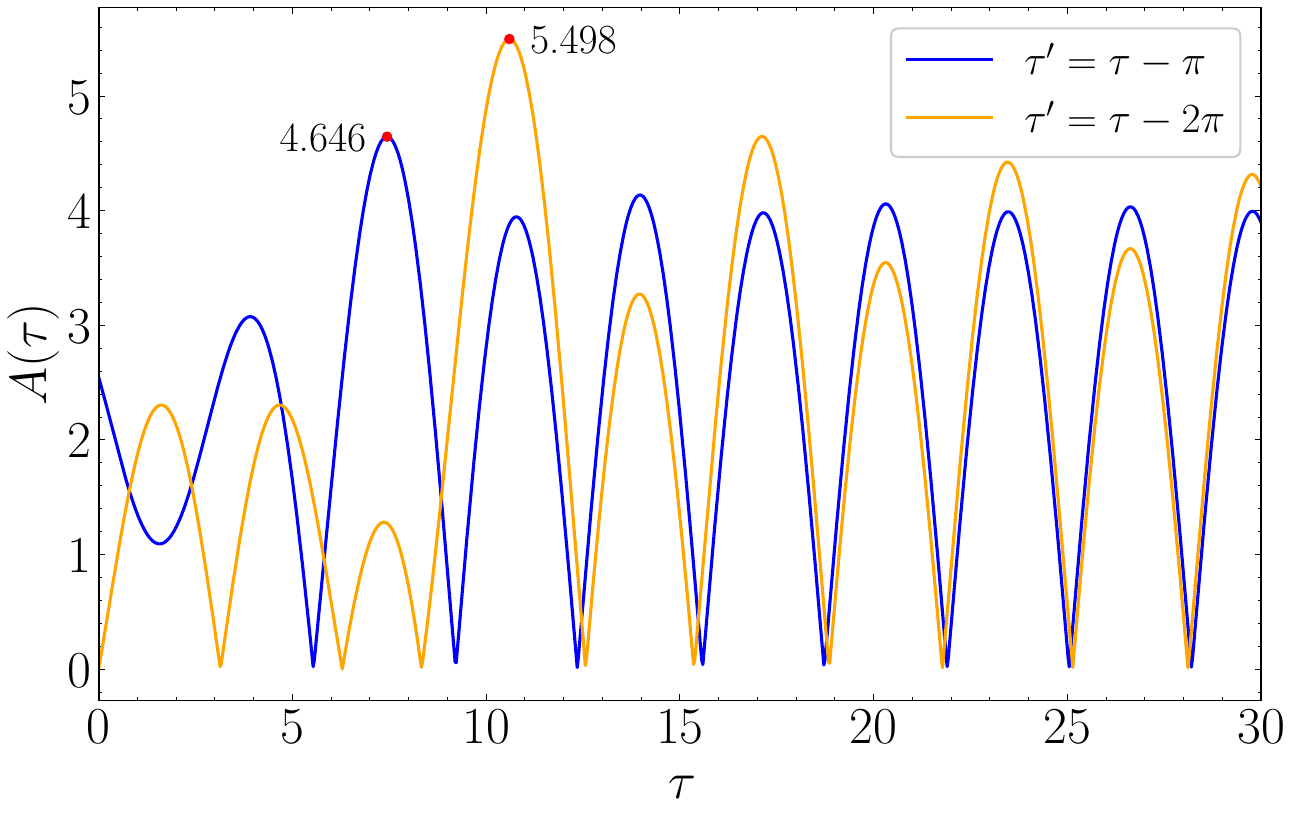}
  \put(3,70){\textbf{(b)}}
\end{overpic}
\caption{(a) Three-dimensional plot of the function $A(\tau, \tau')$ defined in Eq.~\eqref{Eq:A}. Maxima correspond to the maximum kinetic energy gain of electrons in the continuum. (b) Two-dimensional representation under the double recombination conditions ($\tau'=\tau-\pi$) and ($\tau'=\tau-2\pi$). All quantities shown are dimensionless. The corresponding maxima at $4.646U_p$ and $5.498U_p$ are in close agreement with the classical predictions of $4.7U_p$ and $5.5U_p$~\cite{Wang-preprint23, Wang-preprint26, Koval2007}.}
\label{fig:Agraph}
\end{figure}

The following remarks are in order to gain a better understanding of the above expressions and relate them to the results of Ref.~\cite{Wang-preprint23, Wang-preprint26}:
\begin{enumerate}
    \item Ref.~\cite{Wang-preprint23, Wang-preprint26} adopts the classical simple-man's approach to calculate the trajectories leading to simultaneous double recombination. 
    
    \item Two families of classical trajectories were identified (see Fig. 3 of Ref.~\cite{Wang-preprint23, Wang-preprint26}), corresponding to $t'=0$, and either $t''\simeq \pi + O(\epsilon)$ (orange curve in Fig.~\ref{fig:Agraph}(b)) or $t''\simeq 2\pi + O(\epsilon)$ (blue curve in Fig.~\ref{fig:Agraph}(b)).
    
    \item In the present paper, we forcefully impose the condition of simultaneous double recombination. To connect our model with the results of Ref.~\cite{Wang-preprint23, Wang-preprint26}, we incorporate findings from their classical trajectory simulations. This is illustrated in Fig.~\ref{fig:Agraph}. Fig.~\ref{fig:Agraph}(a) shows a three-dimensional plot of the function $A(\tau, \tau')$ over all return times $\tau$ and $\tau'$, while Fig.~\ref{fig:Agraph}(b) presents the cross-sections corresponding to the two established conditions:  

    {\bf Case I:} Trajectories correspond to $\tau-\tau'\simeq \pi$, and classical cutoff scaling $4.7 U_p$ (orange curve in Fig.~\ref{fig:Agraph}(b)).
    
    {\bf Case II:} Trajectories correspond to $\tau-\tau'\simeq 2\pi$, and classical cutoff scaling $5.5 U_p$ (blue curve in Fig.~\ref{fig:Agraph}(b)).

    The maxima of $A(\tau, \tau')$ correspond to the maximum kinetic energy gained by the electrons propagating along different trajectories in the continuum. Wavepacket spreading becomes significant for long trajectories, suppressing the probability of recombination. 
\end{enumerate}

\section{Numerical Results}
\label{sec:V-numerics}
In this section, we present results for a numerical analysis based on the ``quasi''-analytic expression, Eq.~\eqref{Eq:HHG-saddle-TEA}. The calculation reduces to a twofold time-integral over $\tau'$ and $\tau''$, and a fast Fourier transform with respect to the time $t$ to calculate the high-harmonic spectrum generated by intense infrared femtosecond laser pulses. The details of these calculations are discussed in the Appendix. In Fig.~\ref{fig:sfa_spectra}, we present HHG spectra for the two-electron helium system, which is characterised by the binding energies $I_0 = 2.9\,\mathrm{a.u.}$ and $I_1 = 2.0\,\mathrm{a.u.}$ The driving laser pulse is linearly polarised with a wavelength of $\lambda = 800\,\mathrm{nm}$, a total duration of $n_{\text{c}} = 5$ optical cycles, and a carrier-envelope phase of $\phi = \pi/2$. Numerical simulations were performed for two peak laser intensities: $I = 9 \times 10^{14}\,\mathrm{W/cm^2}$ and $I = 3 \times 10^{15}\,\mathrm{W/cm^2}$. Impressively, the harmonic spectrum extends far beyond
the water window, reaching energies as high as $\approx 1.2\,\mathrm{keV}$ (Fig.~\ref{fig:sfa_spectra}(b)).

\begin{figure}[H]
\centering
\begin{overpic}[width=0.48\linewidth]{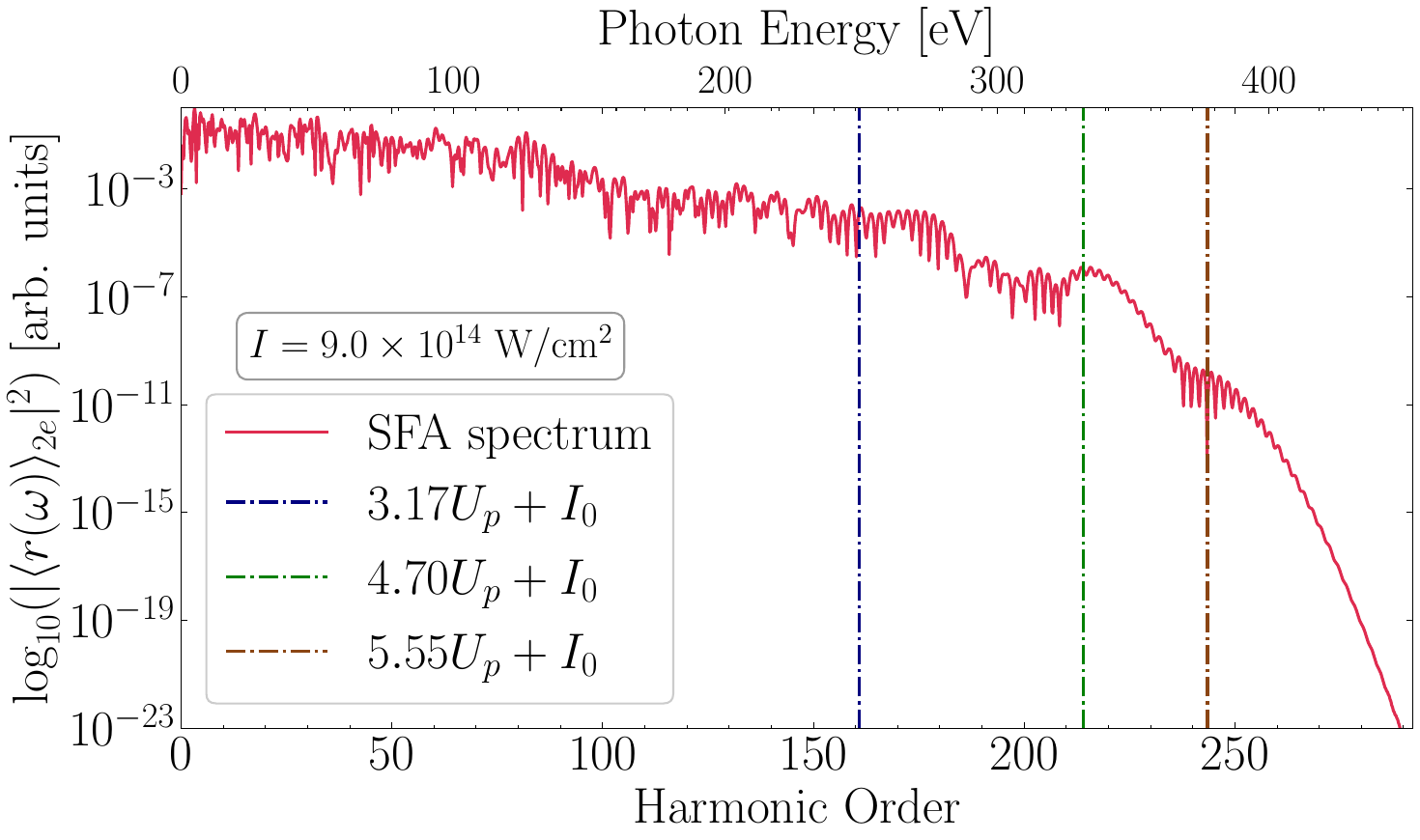}
  \put(3,60){\textbf{(a)}}
\end{overpic}
\hfill
\begin{overpic}[width=0.48\linewidth]{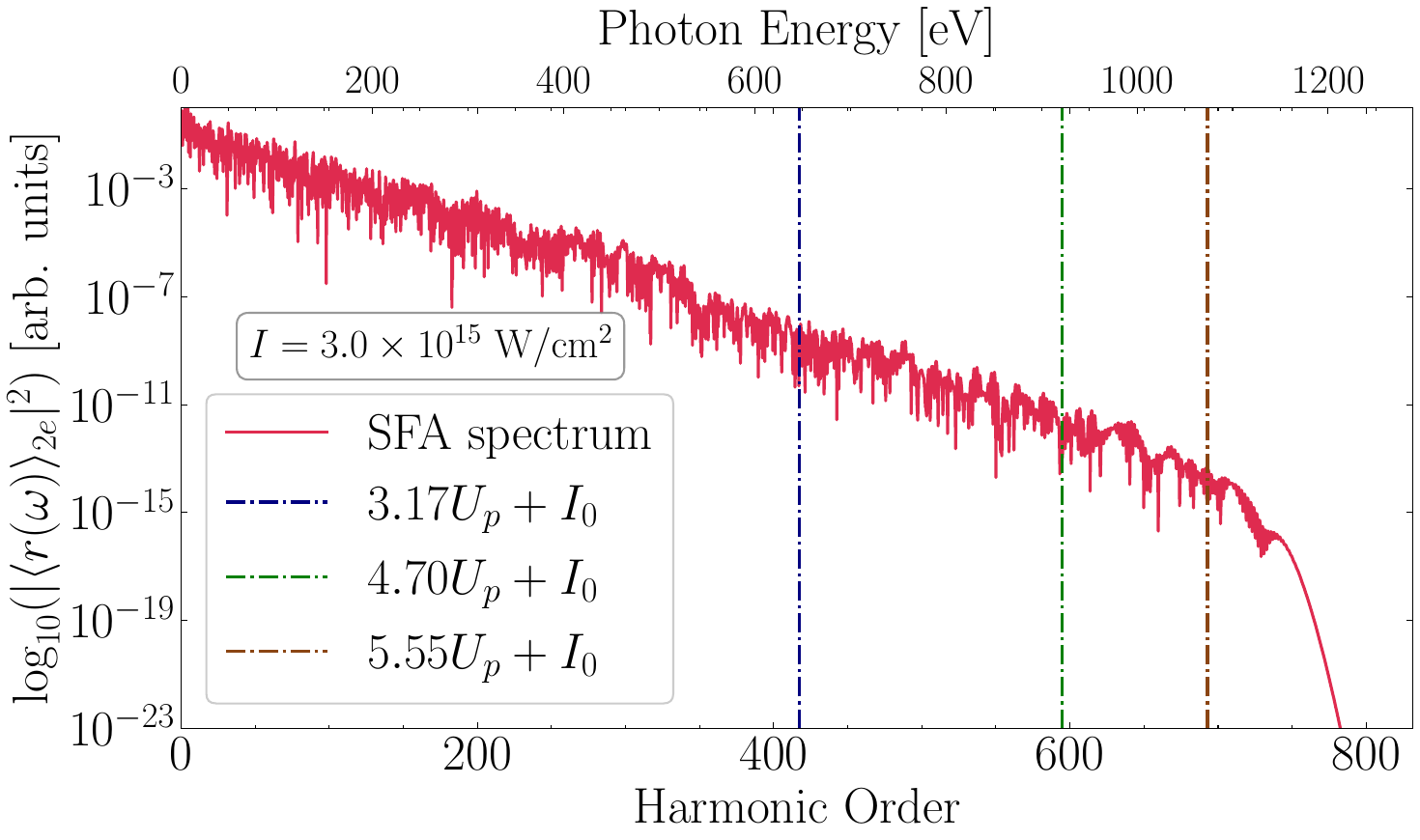}
  \put(3,60){\textbf{(b)}}
\end{overpic}
\caption{HHG spectra calculated using the two-electron SFA for driving laser pulses with peak intensities of (a) $9 \times 10^{14}\,\mathrm{W/cm^2}$ and (b) $3 \times 10^{15}\,\mathrm{W/cm^2}$. In both panels, the vertical dash-dot lines indicate the characteristic cutoff limits given by $(3.17 U_p + I_0)/\omega$, $(4.70 U_p + I_0)/\omega$, and $(5.55 U_p + I_0)/\omega$, demonstrating the extension of the harmonic plateau due to simultaneous double recombination, as observed in Refs.~\cite{Wang-preprint23, Wang-preprint26}.}
\label{fig:sfa_spectra}
\end{figure}

\begin{figure}[H]
\centering

\begin{overpic}[width=0.48\linewidth]{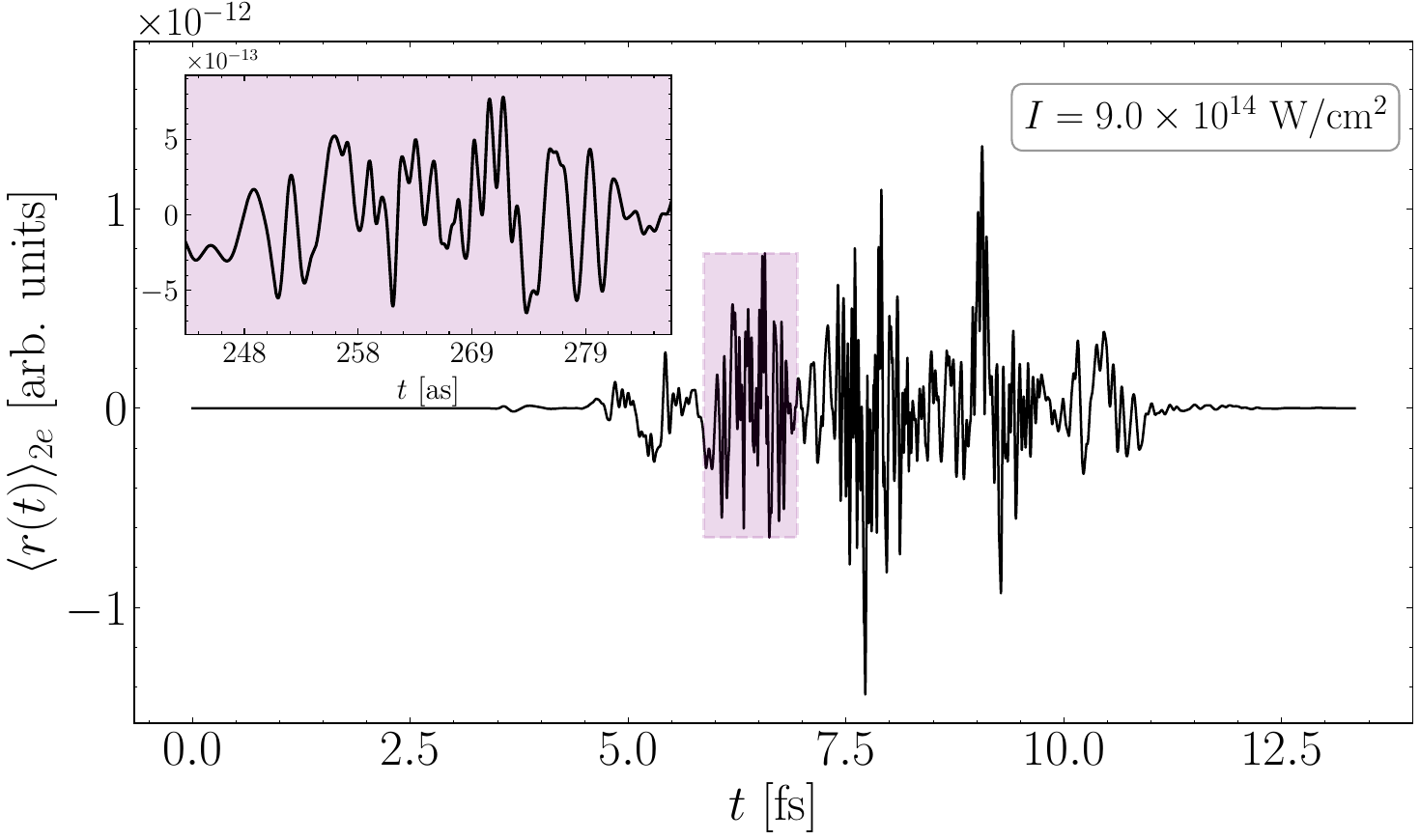}
  \put(3,60){\textbf{(a)}}
\end{overpic}
\hfill
\begin{overpic}[width=0.48\linewidth]{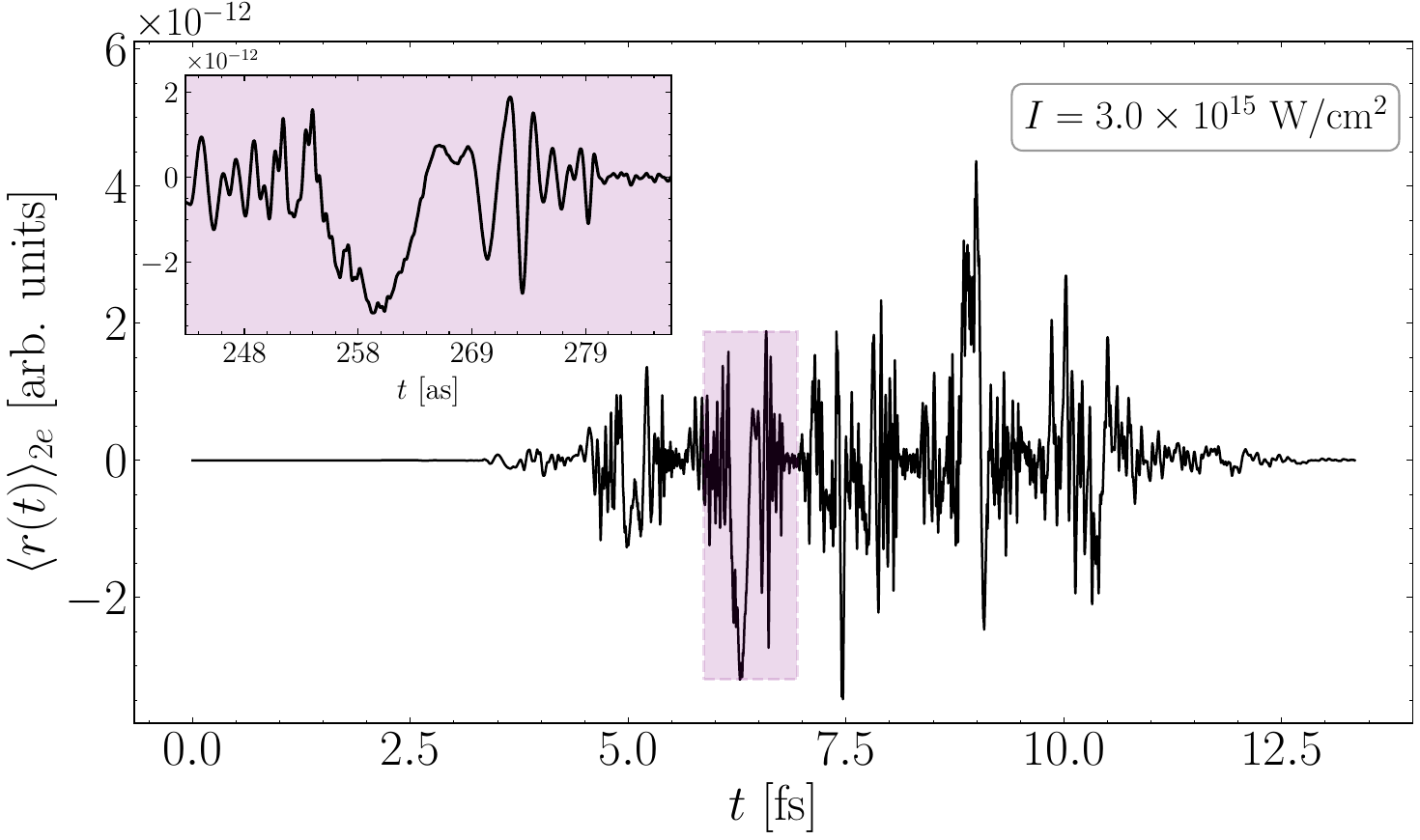}
  \put(3,60){\textbf{(b)}}
\end{overpic}
\caption{Two-electron dipole response within the SFA for driving laser pulses with peak intensities of (a) $9 \times 10^{14}\,\mathrm{W/cm^2}$ and (b) $3 \times 10^{15}\,\mathrm{W/cm^2}$. Both panels include a zoomed-in inset of the shaded region highlighting the strongly oscillatory behaviour.}
\label{fig:sfa_dipole}
\end{figure}

It is also instructive to examine the dipole moment as a function of time (see Fig.~\ref{fig:sfa_dipole}), as this reveals the temporal profile of the emission. The sharp spikes confirm that trajectories at particular recombination times dominate the emission. 

\begin{figure}[H]
    \centering
    \includegraphics[width=0.5\linewidth]{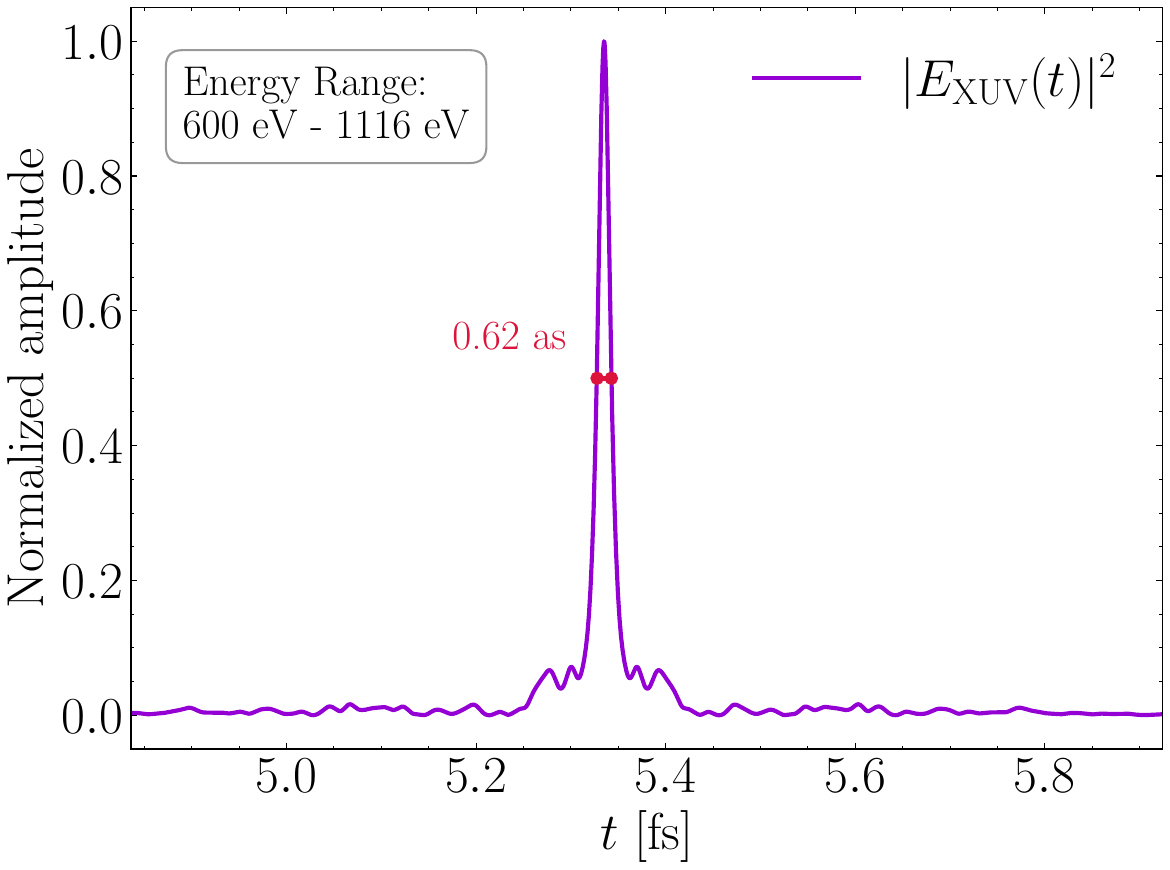}
    \caption{Reconstructed attosecond pulse from the two-electron HHG spectrum shown in Fig.~\ref{fig:sfa_spectra}(b). The reconstruction is performed for a flat-phase model over the energy range from $600\,\mathrm{eV}$ to the cutoff energy $5.5U_{p}+I_{0}\approx 1116\,\mathrm{eV}$.}
    \label{fig:attopulses}
\end{figure}

The present model also facilitates the accurate calculation of the spectral phase distribution across the entire harmonic spectrum, as well as the duration of the pulses formed through the superposition of harmonics. However, this requires extensive analysis involving computationally demanding and time-consuming calculations, which are beyond the scope of the present work. Nevertheless, it is important to note that, assuming a flat spectral phase distribution, the bandwidth of the soft x-ray radiation presented in Fig.~\ref{fig:sfa_spectra}(b) is capable of supporting pulses on sub-attosecond timescales, as shown in Fig.~\ref{fig:attopulses}. This can be experimentally achieved by filtering a set of harmonics near the extended two-electron cutoff using specially designed optical elements or thin metal filters (such as Be, Al) with high transmission across a broad energy range in the soft x-ray regime~\cite{Henke1993}.

\section{Conclusions and Outlook}
\label{sec:VI-conclusion}
In this work, we generalised the ``standard'' SFA framework of Ref.~\cite{Lewenstein1994} to include two-electron effects. Firstly, we reproduced the classical results of Refs.~\cite{Wang-preprint23, Wang-preprint26, Koval2007}, which were derived using the simple man's model. We then extended our analysis beyond constant-envelope laser fields to study short, few-cycle laser pulses. The resulting harmonics extend far beyond the water window, reaching photon energies of up to $\approx1.2\,\mathrm{keV}$ in the soft x-ray regime. Assuming a flat spectral phase distribution, the broad spectral bandwidth can support the generation of sub-attosecond pulses. This analysis provides new insights into the physical and mathematical nature of simultaneous double recombination in HHG, as well as enabling the generation of ultrashort pulses in the water window, which are particularly interesting for applications to core-level spectroscopy and biological imaging. Our results can be generalised in various directions in future work. This includes: I) Extension beyond linear polarisation to elliptically polarised fields, as discussed in Ref.~\cite{Wang-preprint23, Wang-preprint26}; II) Exploration of alternative structured light pulses to enhance the two-electron effects; III) Extension of the present theory to include single-electron HHG and a possible interference between SAE and TAE descriptions. Other relevant competing processes can also be included, e.g. EII, RESI, HHG from $\mathrm{He}^+$ ions, etc.; IV) Comparison of our results with the solutions of the 1D TDSE describing two-electron systems in the spirit of the late J.H. Eberly~\cite{Grobe1992}; V) Investigations of the influence of the discussed two-electron processes on ATI, EII, and RESI mechanisms~\cite{symphony}.

\appendix\section{Numerical methods and approximations}
\label{sec:A-instructions}
Here we outline the analytic tricks employed to reduce Eq.~\eqref{Eq:HHG-td-TEA} to a twofold integral over $\tau$ and $\tau'$. \\

 \noindent{\bf Transition matrix elements}. We follow the standard approach of Ref.~\cite{Lewenstein1994}, approximating relevant states with Gaussians and plane waves in order to estimate the transition matrix elements. The Gaussian wavefunction provides a reasonable approximation because the relevant states in helium are $s$-wave states. The dipole moment, expressed in atomic units ($e=m=\hbar=1$), is given by:
 \begin{equation}
d(p)=i\left(\frac{1}{\pi \alpha}\right)^{3/4}\frac{p}{\alpha}\exp\left(-\frac{p^2}{2\alpha}\right),
 \end{equation}
 
where $\alpha$ corresponds to the binding energy of the state involved in the transition. In particular, the relevant transition dipole moments are: 
 \begin{align}
 d(p, p')&= -\left(\frac{1}{\pi I_0}\right)^{3/2}\frac{pp'}{I_0^2}\exp\left(-\frac{p^2 + (p')^2}{2 I_0}\right), \\
 d_1(p)&=i \left(\frac{1}{\pi I_1}\right)^{3/4}\frac{p}{I_1}\exp\left(-\frac{p^2}{2I_1}\right).
 \end{align}
 
Here we assume the dipole moment associated with simultaneous double recombination can be factorised as $d(p,p') = d_0(p)d_0(p')$. This assumes separable continuum states $|p,p'\rangle = |p\rangle|p'\rangle$, neglecting electron-electron correlations during continuum propagation. The resulting expression represents the leading-order approximation within the SFA framework. \\

 \noindent{\bf Laser field}. We start by defining the electric field of the laser pulse as follows, choosing a convenient form for the temporal envelope to simplify calculations: 
\begin{equation} \label{Eq:E(t)}
    E(t)= E_0 \sin^2\left(\frac{\omega t}{2 n_c}\right)\sin(\omega t-\phi).
\end{equation}
Here $\phi$ is the carrier-envelope phase (CEP), while $n_c$ is the number of cycles in the pulse, which ranges from $2-10$ in typical applications. We abandon vector notation since the field is linearly polarised along the $x$-axis and all nontrivial vector components are oriented in the $x$-direction. Clearly, $E(0)=E(t_f = 2\pi n_c/\omega)= 0$. The vector potential is related to the field by $E(t)=-\frac{1}{c}\frac{\partial A(t)}{\partial t}$, so it is simple to solve analytically for $A(t)$:
\begin{equation} \label{Eq:A(t)}
    \frac{eA(t)}{c}= \frac{eE_0}{2\omega} \cos(\omega t-\phi) - \frac{eE_0}{4\omega} \frac{\cos(\omega t-\phi +\omega t/n_c)}{1+1/n_c} -  \frac{eE_0}{4\omega} \frac{\cos(\omega t-\phi -\omega t/n_c)}{1-1/n_c}.
\end{equation} 
Note also that $A(0)=A(t_f)=0$, so the canonical and kinetic momenta coincide at the initial and final times.

We can further simplify notation by adopting atomic units and expressing the relevant energies (the ponderomotive energy $U_p$ and the ionisation potentials $I_0$ and $I_1$) in units of laser photon energy, $\hbar\omega$. As in Section~\ref{sec:IV-Saddle}, we rescale $t \to \omega t$, and additionally define the dimensionless parameter $\Delta = 1/(2n_c)$. 

 This results in the scaled vector potential:
\begin{equation}
    \frac{eA(t)}{c}= \sqrt{U_p}\tilde A(t),    
\end{equation} 
where
\begin{equation} \label{Eq:A_tilde(t)}
\tilde A(t) = \cos(t-\phi) - \frac{1}{2} \frac{\cos(t-\phi +2\Delta t)}{1+2\Delta} -  \frac{1}{2} \frac{\cos(t-\phi -2\Delta t)}{1-2\Delta}.
\end{equation}\\

\noindent{\bf Saddle-point momenta}. The saddle-point values of the canonical momenta, corresponding to the stationary phase of the quasi-classical action, are obtained from Eqs.~\eqref{Eq:saddle_p1} and~\eqref{Eq:saddle_p2}: 
\begin{align}
p_{\rm st}(t, \tau)&= \frac{e}{c\tau}\int_{t-\tau}^t d{\tilde t} \; A({\tilde t}) = \frac{\sqrt{U_p}}{\tau}\int_{t-\tau}^t d{\tilde t} \; \tilde A({\tilde t}) , \label{Eq:pst1}\\
p_{\rm st}(t, \tau')&=\frac{e}{c\tau'}\int_{t-\tau'}^t d{\tilde t} \; A({\tilde t}) = \frac{\sqrt{U_p}}{\tau'}\int_{t-\tau'}^t d{\tilde t} \; \tilde A({\tilde t}) \label{Eq:pst2}.
\end{align}
 The above integrals can be evaluated analytically:

\begin{align}\
 \frac{\sqrt{U_p}}{\tau}\int_{t-\tau}^t d{\tilde t} \; \tilde A({\tilde t}) &= \frac{\sqrt{U_p}}{\tau}\left[\sin(\tilde t-\phi) - \frac{1}{2} \frac{\sin(\tilde t-\phi +2\Delta \tilde t)}{(1+2\Delta)^2} - \frac{1}{2} \frac{\sin(\tilde t-\phi -2\Delta \tilde t)}{(1-2\Delta)^2}\right]\Big|_{t-\tau}^t, \\
 \frac{\sqrt{U_p}}{\tau'}\int_{t-\tau'}^t d{\tilde t} \; \tilde A({\tilde t}) &= \frac{\sqrt{U_p}}{\tau'}\left[\sin(\tilde t-\phi) - \frac{1}{2} \frac{\sin(\tilde t-\phi +2\Delta \tilde t)}{(1+2\Delta)^2} - \frac{1}{2} \frac{\sin(\tilde t-\phi -2\Delta \tilde t)}{(1-2\Delta)^2}\right]\Big|_{t-\tau'}^t.
\end{align}

 These solutions correspond to the ``Broad Gaussian Limit'' (BGL) in Ref.~\cite{Lewenstein1994}, where the saddle-point values are fully determined by the stationary phase of the quasi-classical action without the need for corrections due to the variations in the dipole matrix elements. Therefore, we can greatly simplify the double integral over momenta by replacing the momenta everywhere with the saddle-point solutions, and introducing the following factors to account for Gaussian fluctuations around the saddle-points in three dimensions:
 \begin{equation} 
\left(\frac{\pi}{1/(I_0-I_1) + i\tau/2}\right)^{3/2}; \: \: \: \left(\frac{\pi}{1/I_1 + i\tau'/2}\right)^{3/2}.
 \end{equation}
 In the BGL, these factors are regularised by an infinitesimal parameter $\epsilon$. However, here we generalise the BGL by accounting for regularisation from the actual fluctuations in the dipole transition matrix elements.\\

Using Eqs.~\eqref{Eq:pst1} and~\eqref{Eq:pst2}, the quasi-classical action in Eq.~\eqref{Eq:s_eff} simplifies to the following expression at the saddle points:
\begin{equation}
    \begin{aligned}
        {S}_{\rm eff}(p, p', t', t'',t) =& (I_0-I_1)\tau 
+ \int_{t'}^{t}d{\tilde t}
\left[
\frac{1}{2m}(p-eA({\tilde t})/c)^2\right]  + I_1\tau' +\int_{t'}^{t}d{\tilde t}
\left[\frac{1}{2m}(p'-eA({\tilde t})/c)^2
\right]  \\ &- \hbar (2K+1)\omega t \\
 =& (I_0-I_1)\tau 
- \frac{p_{st}(t,\tau)^2 \tau}{2} + \frac{U_p}{2}\int_{t-\tau}^{t}d{\tilde t}
({\tilde A}({\tilde t}))^2 + I_1\tau' - \frac{p_{st}(t,\tau')^2 \tau'}{2} \\&+ \frac{U_p}{2}\int_{t-\tau'}^{t}d{\tilde t}
({\tilde A}({\tilde t}))^2 - (2K+1) t ,
    \end{aligned}
    \label{Eq:s_eff_num}
\end{equation}

Integrals of the form $\int_{t-\tau}^{t}d{\tilde t}
{(\tilde A}({\tilde t}))^2$ in Eq.~\eqref{Eq:s_eff_num} can be evaluated in closed form in a trivial but tedious calculation: 
\begin{equation}
    \begin{aligned}
        \int_{t-\tau}^{t}d{\tilde t}
(\tilde A(\tilde t))^2 =& \left[ \frac{\tilde t}{2} + \frac{\sin{(2{\tilde t}-2\phi)}}{4} + \frac{\tilde t}{8(1+2\Delta)^2} + \frac{\sin{(2\tilde t-2\phi+4\Delta \tilde t)}}{16(1+2\Delta)^3} + \frac{\tilde t}{8(1-2\Delta)^2} \right. \\
 &  \left. +
\frac{\sin{(2\tilde t-2\phi-4\Delta \tilde t)}}{16(1-2\Delta)^3} - \frac{\sin{(2\tilde t-2\phi+2\Delta \tilde t)}}{4(1+2\Delta)(1+\Delta)} - \frac{\sin{(2\Delta \tilde t)}}{4\Delta(1+2\Delta)} - \frac{\sin{(2\tilde t -2\phi-2\Delta \tilde t)}}{4(1-2\Delta)(1-\Delta)} \right. \\
& \left. - \frac{\sin{(2\Delta \tilde t)}}{4\Delta(1-2\Delta)} + \frac{\sin{(2\tilde t - 2\phi)}}{8(1-4 \Delta^2)} + \frac{\sin{(4\Delta \tilde t)}}{16\Delta(1-4\Delta^2)} \right] \Big|_{t-\tau}^t .
    \end{aligned}
\end{equation}

\noindent{\bf Dipole moment}. It is now possible to numerically evaluate the dipole moment in Eq.~\eqref{Eq:HHG-td-TEA}, where, for simplicity, we consider the  weak depletion limit, and set $a(t)=a^*(t)=1$:
\begin{equation}
    \begin{aligned}
        \langle r(t)\rangle_{2e} 
& =
\mathrm{Re}\left[
\left(\frac{\textit{i}}{\hbar}\right)^2 
\int_0^{t}\textit{d} 
\tau \: \int_{t-\tau}^{t}\textit{d} 
\tau' \, \left[\left(\frac{\pi}{1/(I_0-I_1) + i\tau/2}\right) \left(\frac{\pi}{1/I_1 + i\tau'/2}\right)\right]^{3/2} \right. \\
& \left. \times
d^*\left(p_{st}(t,\tau)-\sqrt{U_p}\tilde{A}(t), p_{st}(t,\tau')-\sqrt{U_p} \tilde A(t)\right)
 \; E(t-\tau') \; E(t-\tau) \; d_0(p_{st}(t,\tau)-\sqrt{U_p}\tilde{A}(t-\tau)) \right. \\ 
 & \left. \times d_1(p_{st}(t,\tau')-\sqrt{U_p}\tilde{A}(t-\tau')) \;\exp{\left(
  {- i \:
  {S}(p_{st}(t,\tau), t-\tau, t-\tau')/\hbar }\right)}
  \; \exp{\left(
  -{\textit{i} \:
  {S}(p_{st}(t,\tau}), p_{st}(t,\tau'),t-\tau',t)/\hbar 
  \right)} \right.  \\
&\left. + (p \leftrightarrow p', \tau \leftrightarrow \tau') \right] .
    \end{aligned}
    \label{Eq:HHG-saddle-TEA}
\end{equation}
To numerically evaluate the two-electron dipole response given by Eq.~\eqref{Eq:HHG-saddle-TEA}, we implement a non-adaptive spectral integration scheme. For each of the $N_t = 2^{15}$ sampled temporal grid points, the dipole moment involves a double integral over the ordered, coupled excursion times within the triangular domain defined by $0 < \tau' < \tau < t$. We map this triangular domain to a unit square $[0, 1] \times [0, 1]$ via the linear transformation $\tau' = t u$ and $\tau = \tau' + (t - \tau') v$. The transformed integrals over the independent variables $u$ and $v$ are then evaluated using a two-dimensional Gauss-Legendre quadrature with $N_{\text{quad}} = 96$ nodes.

Finally, we calculate the HHG spectrum $\tilde r(\Omega)$ by evaluating the Fourier transform of $\langle r(t)\rangle_{2e}$. This can be performed efficiently with a standard  FFT algorithm:

\begin{equation} 
\tilde r(\Omega) = \frac {1}{2\pi}\int_0^{t_f}dt \langle r(t)\rangle_{2e} \exp(-i\Omega t) .
\end{equation}

\acknowledgments{ 
We thank Jens Biegert, Phil Bucksbaum, Carla Faria, Arti Gaharwar, Lidija Petrovic, and Emilio Pisanty for stimulating discussions.  

ICFO-QOT group acknowledges support from: European Research Council AdG NOQIA; MCIN/AEI (PGC2018-0910.13039/501100011033,  CEX2019-000910-S/10.13039/501100011033, Plan National FIDEUA PID2019-106901GB-I00, Plan National STAMEENA PID2022-139099NB, I00, project funded by MCIN / AEI / 10.13039 / 501100011033 and by the “European Union NextGenerationEU/PRTR" (PRTR-C17.I1), FPI); QUANTERA DYNAMITE PCI2022-132919, QuantERA II Programme co-funded by European Union’s Horizon 2020 program under Grant Agreement No 101017733; Ministry for Digital Transformation and of Civil Service of the Spanish Government through the QUANTUM ENIA project call - Quantum Spain project, and by the European Union through the Recovery, Transformation and Resilience Plan - NextGenerationEU within the framework of the Digital Spain 2026 Agenda; Fundació Cellex; Fundació Mir-Puig; Generalitat de Catalunya (European Social Fund FEDER and CERCA program; Barcelona Supercomputing Center MareNostrum (FI-2023-3-0024); Funded by the European Union. Views and opinions expressed are however those of the author(s) only and do not necessarily reflect those of the European Union, European Commission, European ClimateInfrastructure and Environment Executive Agency (CINEA), or any other granting authority.  Neither the European Union nor any granting authority can be held responsible for them (HORIZON-CL4-2022-QUANTUM-02-SGA  PASQuanS2.1, 101113690, EU Horizon 2020 FET-OPEN OPTOlogic, Grant No 899794, QU-ATTO, 101168628),  EU Horizon Europe Program (This project has received funding from the European Union’s Horizon Europe research and innovation program under grant agreement No 101080086 NeQSTGrant Agreement 101080086 — NeQST); ICFO Internal “QuantumGaudi” project. 

A. M and M. F. C. acknowledge support by the Quantum Science and Technology-National Science and Technology Major Project (Grant No. 2025ZD0301000),  the National Key Research and Development Program of China (Grant No. 2023YFA1407100), the Guangdong Province Science and Technology Major Project (Future functional materials under extreme conditions - 2021B0301030005) and the National Natural Science Foundation of China (Grant No. 12574092). 

P.T. at FORTH acknowledges the European Union’s HORIZON-MSCA-2023-DN-01 project QU-ATTO under the Marie Skłodowska-Curie grant agreement No 101168628 and ELI–ALPS. The ELI-ALPS project (GINOP-2.3.6-15-2015-00001) is supported by the EU and co-financed by the European Regional Development Fund. 

J. R.-D. acknowledges funding from UK Engineering and Physical Sciences Research Council (EPSRC) Funding, Grant UKRI2300 - Attosecond Photoelectron Imaging with Quantum Light (APIQuL).
}

\appendix

\bibliographystyle{unsrt} 
\bibliography{references}{}

\end{document}